\documentclass[aps,prb,twocolumn,showpacs,preprintnumbers]{revtex4}
\usepackage{bm,hyperref}
\usepackage{amsmath}
\usepackage{graphicx}
\usepackage{soul}
\usepackage{amsfonts}
\usepackage{amssymb}
\usepackage{dcolumn}
\usepackage{amsmath}
\usepackage{esint}

\def\-#1{{\bf #1}}

\begin{document}
\title{All-dielectric reciprocal bianisotropic nanoparticles}
\author{Rasoul Alaee{$^{*,1}$}, Mohammad Albooyeh{$^{2}$}, Aso Rahimzadegan{$^{1}$}, Mohammad S. Mirmoosa{$^{2}$}, Yuri S. Kivshar{$^{3}$}, and Carsten Rockstuhl{$^{1,4}$}}
\address{
{$^{1}$}Institute of Theoretical Solid State Physics, Karlsruhe Institute of Technology, Karlsruhe 76131, Germany\\
{$^{2}$}Department of Radio Science and Engineering, School of Electrical Engineering, Aalto University, Aalto, Finland\\
{$^{3}$}Nonlinear Physics Centre, Research School of Physics and Engineering, Australian National University,
Canberra ACT 0200, Australia\\
{$^{4}$}Institute of Nanotechnology, Karlsruhe Institute of Technology, Karlsruhe 76021, Germany\\
$^*$Corresponding author: rasoul.khanghah@kit.edu
}

\begin{abstract}
The study of high-index dielectric nanoparticles currently attracts a lot of attention. They do not suffer from absorption but promise to provide control on the properties of light comparable to plasmonic nanoparticles. To further advance the field, it is important to identify versatile dielectric nanoparticles with unconventional properties. Here, we show that breaking the symmetry of an all-dielectric nanoparticle leads to a geometrically tunable magneto-electric coupling, i.e. an omega-type bianisotropy. The suggested nanoparticle exhibits different backscatterings and, as an interesting consequence, different optical scattering forces for opposite illumination directions. An array of such nanoparticles provides different reflection phases when illuminated from opposite directions. With a proper geometrical tuning, this bianisotropic nanoparticle is capable of providing a $2\pi$ phase change in the reflection spectrum while possessing a rather large and constant amplitude. This allows creating reflectarrays with near-perfect transmission out of the resonance band due to the absence of an usually employed metallic screen.
\end{abstract}
\maketitle
Nanophotonics has attracted enormous research interests due to its potential to control light-matter interaction at the nanoscale \cite{Novotny:06,Maier:07,Gramotnev:10,Atwater:10}. Being usually connected with a strong light confinement in metal-dielectric and plasmonic structures,  nanophotonics offers remarkable opportunities due to the local field enhancement~\cite{Schuller:10, Albooyeh:12, Alaee:13}. However, applications of plasmonic nanophotonics are often limited by Ohmic losses at optical frequencies \cite{Boltasseva:11,Tassin:12,Filonov:12,Moitra:13,Alaee:OL15}. An alternative strategy in nanophotonics is to use high-index dielectric materials for building blocks that control light-mater interaction \cite{Ahmadi:08,Evlyukhin:10,Evlyukhin:12}. Dielectric nanoparticles have attracted a considerable attention due to their appealing applications \cite{Krasnok:12,Yang:14,Krasnok:14,Brongersma:14,Lin:14,Staude:15,Jain:14}.

\begin{figure}
\centering
\includegraphics[width=0.48\textwidth]{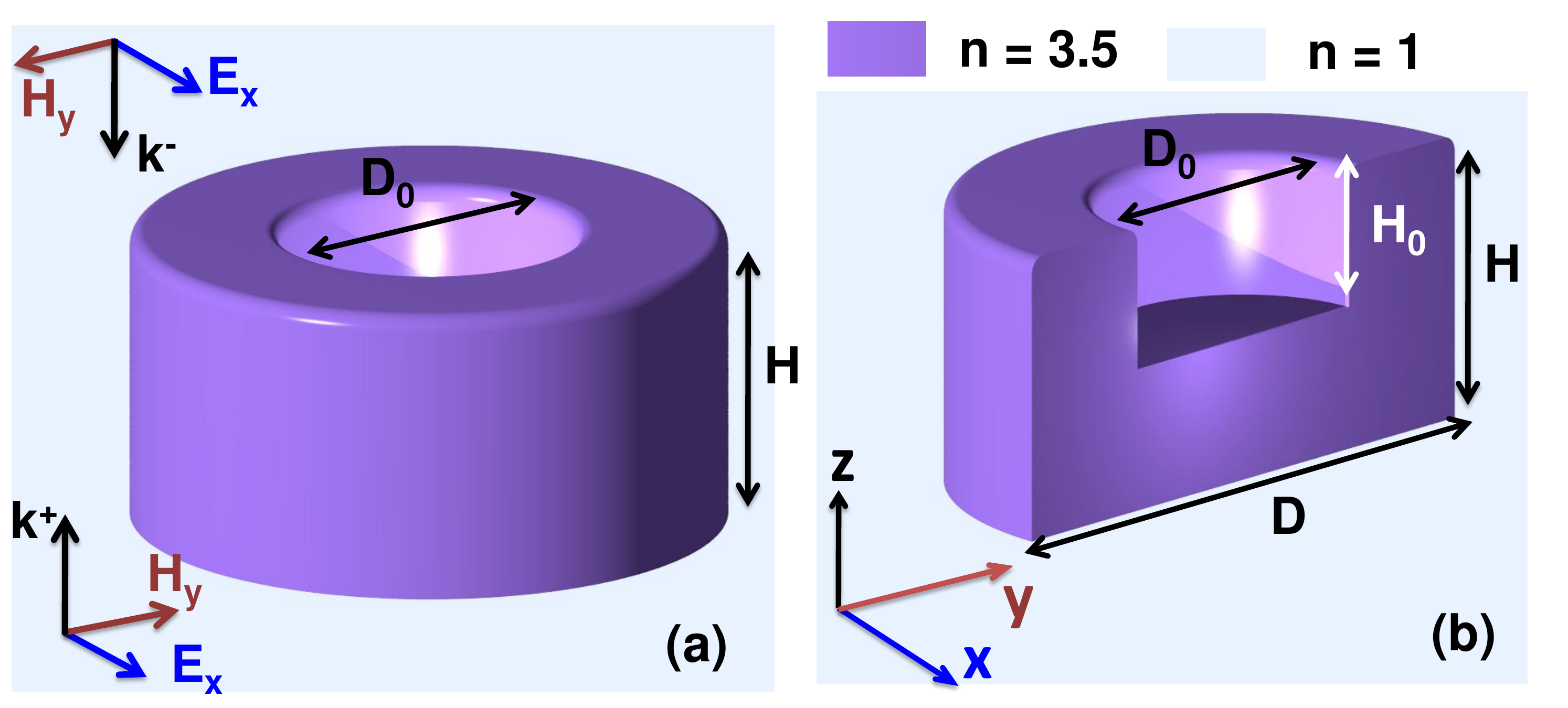}
\caption{(a) Schematic of the proposed bianisotropic high-index dielectric nanoparticle. (b) A cut in the $yz$-plane of the proposed nanonatenna. The dielectric cylinder has a diameter $D=300$ nm and height $H=150$ nm with a cylindrical hole inside it that has a diameter $D_{0}=150$ nm and height $H_{0}=10-140$ nm. We assumed that the refractive index of the material from which the cylinder is made is $n =$ 3.5. The ambient material shall be air with a refractive index of unity.
}\label{fig:figGeo}
\end{figure}

For these nanoparticles to be versatile, usually the multipolar composition of their scattered field is tailored. A high-index dielectric nanoparticle that supports both electric and magnetic resonant responses where the exact composition of the two contributions can be tuned by means of geometrical modifications has been recently suggested~\cite{Schuller:07,Kuznetsov:12,Staude:13}. Interesting optical features such as directional scattering pattern with zero backscattering or an optical pulling force can be achieved by proper tuning of these responses~\cite{Kerker:83,Chen:11,Saenz:11,Puyalto:13, Steven:13}. Nanoparticles with higher multipolar responses, e.g. electric and magnetic quadrupoles may provide an even more control on the scattering properties~\cite{Liu:14,Mirzaei:15} when compared to nanoparticles with only dipolar responses. However, they require more complicated theoretical considerations and a range of dissimilar materials \cite{Mirzaei:15}. Therefore, it is a challenge to bring more control on the scattering features of nanoparticles without adding such complexities. Here, we use \textit{bianisotropy} to control the scattering properties of high-index dielectric nanoparticles in the context of \textit{dipolar} responses~\cite{Serdyukov:01,Alaee:15}.
Note that we are only interested in the reciprocal omega-type bianisotropy and all the other types of bianisotropy (i.e. chiral, Tellegen, and “moving”) are absent \cite{Serdyukov:01,priou:12}. We show that an omega-type bianisotropic response can be achieved by breaking the symmetry of a cylindrical nanoparticle, i.e. a high-index dielectric cylinder with partially drilled cylindrical air hole [see Figs.~\ref{fig:figGeo}(a) and ~\ref{fig:figGeo}(b))].

A reciprocal omega-type bianisotropic nanoparticle supports electric and magnetic dipole moments with different strength when illuminated by a plane wave in forward or backward directions [see Fig.~\ref{fig:figGeo}(a)]. In particular, the bianisotropy causes significant difference in backscattering responses, and consequently in the optical scattering forces exerted on the proposed nanoparticle for opposite illumination directions. The observed magneto-electric coupling can be tuned by controlling geometric parameters of the nanoparticle. We demonstrate that an infinite periodic planar array of such lossless nanoparticles (a metasurface), provides different \textit{resonant} reflection phases for different illumination directions. This is due to the proposed magneto-electric coupling, and it is certainly not possible with only an electric and/or magnetic response. We also demonstrate that the investigated nanoparticle is an excellent candidate for unit-cells of reflectarrays. Indeed, this nanoparticle in its reflection resonance band, provides a $2\pi$ phase change for a proper geometrically tuning while it maintains a considerable large and constant amplitude, similar to Huygens' metasurfaces \cite{Decker:15}. Moreover, since there is no metallic ground plate in the proposed reflectarray, it will be transparent out of its resonance band.


\section{Bianisotropic nanoparticles}

The geometry of the proposed bianisotropic high-index dielectric nanoparticle is shown in Figs.~\ref{fig:figGeo}(a) and ~\ref{fig:figGeo}(b). \begin{figure}
\centering
\includegraphics[width=0.49\textwidth]{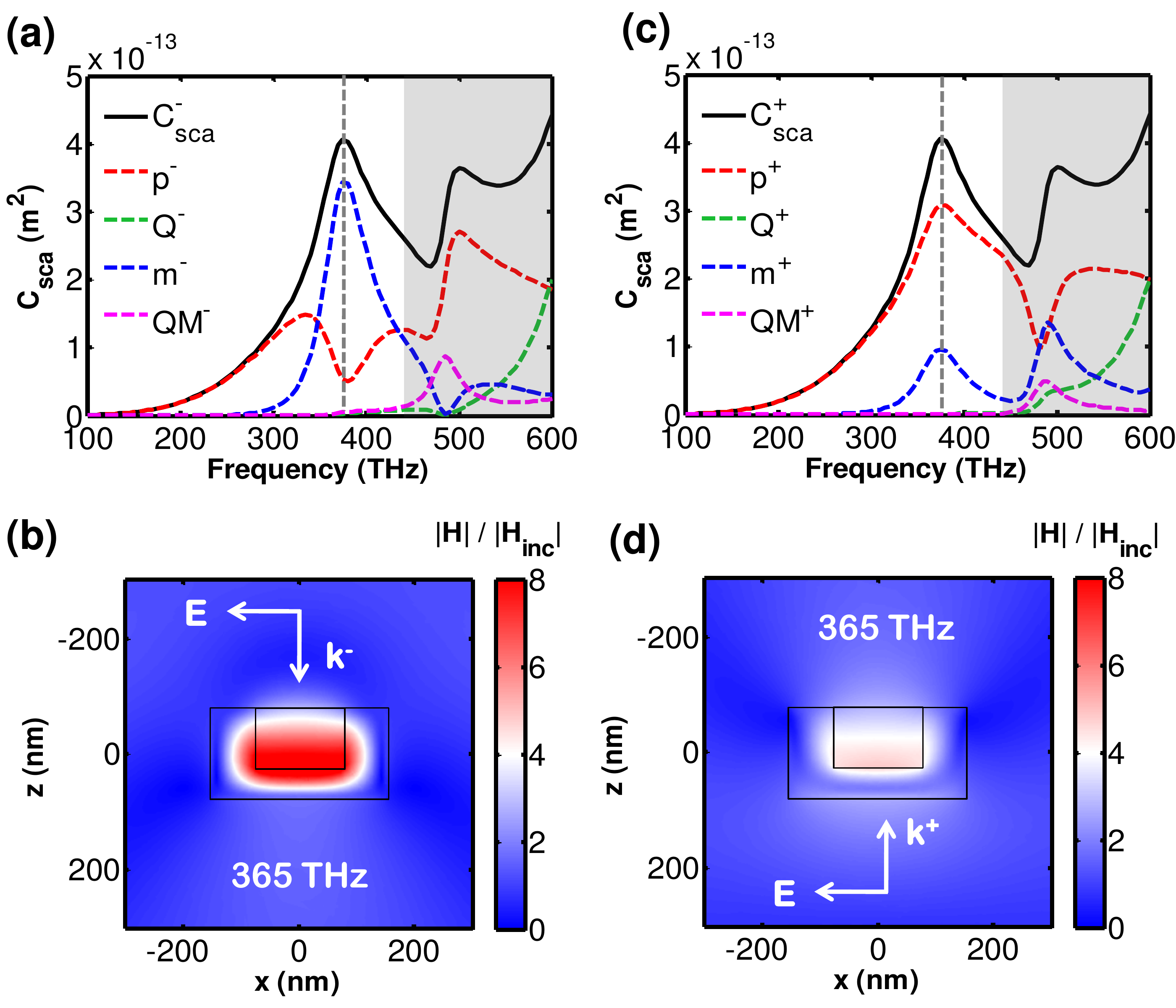}
\caption{(a) and (c) Total scattering cross sections $C_\mathrm{sca}^\pm$ and
contributions from different Cartesian multipole moments
as a function of frequency for forward ($+$) and backward ($-$) illumination directions; electric dipole moment $p^\pm$ (red dashed line),
magnetic dipole moment $m^\pm$ (blue dashed line), electric quadrupole moment $Q^\pm$ (green dashed line) and magnetic quadrupole moment $QM^\pm$ (purple dashed line). The height of the air hole is $H_{0}=100$ nm. (b) and (d) normalized magnetic field distributions at $365\,$THz [gray dashed line in (a) and (c)] for both opposite illumination directions.
}\label{fig:figSCS}
\end{figure}
The suggested nanoparticle is not symmetric with respect to the forward and backward illumination directions [see Fig. \ref{fig:figGeo}(a)]. Material and geometric parameters of the nanoparticle are given in Fig.~\ref{fig:figGeo}. Using the multipole expansion of the scattered field~\cite{muhlig:11}, we compute the total scattering cross section and the contribution of each multipole moment when the nanoparticle is illuminated by a plane wave in the forward and backward directions [see Figs.~\ref{fig:figSCS}(a) and \ref{fig:figSCS}(c)]. The higher order multipole moments, i.e. electric quadrupole \textrm{Q} and magnetic quadrupole \textrm{QM} moments are negligible below 450 THz [the frequency range above 450 THz, signifying the upper frequency for the applicability of the dipole approximation, is marked by the gray shadowed areas in Figs.~\ref{fig:figSCS}(a) and \ref{fig:figSCS}(c)]. In the following, we only concentrate on this frequency range (i.e. lower than 450 THz where only electric and magnetic dipole moments are dominant). The corresponding electric $p_{\textrm{x}}^{\pm}$ and magnetic $m_{\textrm{y}}^{\pm}$ dipole moments of the nanoparticle for both illumination directions [see Fig.~\ref{fig:figGeo}] are given by:
\begin{eqnarray}
\frac{p_{x}^{\pm}}{\epsilon_{0}} & = & \alpha_{\textrm{ee}}E_{x}^{\textrm{inc}}\pm\alpha_{\textrm{em}}Z_{0}H_{y}^{\textrm{inc}},\label{pm1}\\
Z_{0}m_{y}^{\pm} & = & \alpha_{\textrm{me}}E_{x}^{\textrm{inc}}\pm\alpha_{\textrm{mm}}Z_{0}H_{y}^{\textrm{inc}},\label{pm2}
\end{eqnarray}
where $\epsilon_0$  is the free space permittivity and $Z_0$ is the characteristic impedance of free space. $\alpha_{\textrm{ee}}$, $\alpha_{\textrm{em}}$, $\alpha_{\textrm{me}}$, $\alpha_{\textrm{mm}}$ are individual polarizabilities of the nanoparticle. They represent electric, magneto-electric, electro-magnetic, and magnetic couplings in the proposed nanoparticle, respectively. That is, the interactions between the electric and magnetic fields and polarizations \cite{Serdyukov:01,albooyeh2011substrate}. Notice that in order to distinguish the illumination direction of the nanoparticle, we use the ${\pm}$ sign where the plus/minus sign corresponds to the propagation direction of the incident plane waves in forward/backward direction (the time dependency is assumed to be $e^{-i\omega t}$).
In order to explain the underlying physical mechanisms of the scattering response of the bianisotropic nanoparticle, we start with the definition of extinction cross section $C_{\textrm{ext}}^{\pm}$ for a bianisotropic nanoparticle in dipole approximation for both illumination directions, i.e.~\cite{Huffman:98,Belov:03,Novotny:06}

\begin{eqnarray}\label{C_Ex} \displaystyle
C_{\textrm{ext}}^{\pm} & = & \frac{P_{\textrm{ext}}^{\pm}}{I_{0}}\nonumber\\
 & = & \frac{-\frac{1}{2}\textrm{Re}\oiint_{S}\left(\mathbf{E}_{\textrm{inc}}\times\mathbf{H}_{\textrm{sca}}^{\pm*}+\mathbf{E}_{\textrm{sca}}^{\pm}\times\mathbf{H}_{\textrm{inc}}^{*}\right)\cdot\mathbf{n}ds}{I_{0}}\nonumber\\
 & = & k\textrm{Im}\left(\alpha_{\textrm{ee}}\pm\alpha_{\textrm{em}}\pm\alpha_{\textrm{me}}+\alpha_{\textrm{mm}}\right),
\end{eqnarray}
where $I_{0}=|\mathbf{E}_{\textrm{inc}}|/2Z_{0}$ is the intensity of the incident plane wave. $|\mathbf{E}_{\textrm{inc}}|$  and $|\mathbf{E}_{\textrm{sca}}^{\pm}|$ are the amplitude of the incident and scattered electric fields for both illumination directions ($\pm$). $P_{\textrm{ext}}^{\pm}$ is the extracted power by the nanoparticle, $k=\omega/c$ is the wavenumber for an angular frequency $\omega$. For a reciprocal nanoparticle \cite{Serdyukov:01}, i.e. $\alpha_{\textrm{em}}=-\alpha_{\textrm{me}}$, we can conclude that the extinction cross section for both illuminations are identical, i.e. $C_{\textrm{ext}}=C_{\textrm{ext}}^{+}=C_{\textrm{ext}}^{-}$ and is given by\begin{eqnarray}\label{C_EX2}
C_{\textrm{ext}} & = & k\textrm{Im}\left(\alpha_{\textrm{ee}}+\alpha_{\textrm{mm}}\right),
\end{eqnarray}
On the other hand, we know for the proposed nanoparticle that the scattering cross sections for both illuminations can be written as \cite{Huffman:98,Belov:03,Novotny:06}
\begin{eqnarray}\label{C_sca}
C_{\textrm{sca}}^{\pm} & = & \frac{P_{\textrm{sca}}^{\pm}}{I_{0}}\nonumber\\
 & = & \frac{\frac{1}{2}\textrm{Re}\oiint_{S}\left(\mathbf{E}_{\textrm{sca}}^{\pm}\times\mathbf{H}_{\textrm{sca}}^{\pm{}^{*}}\right)\cdot\mathbf{n}ds}{I_{0}}\nonumber\\
 & = & \frac{k^{4}}{6\pi}\left(\left|\alpha_{\textrm{ee}}\pm\alpha_{\textrm{em}}\right|^{2}+\left|\alpha_{\textrm{me}}\pm\alpha_{\textrm{mm}}\right|^{2}\right),
\end{eqnarray}
where, $P_{\textrm{sca}}^{\pm}$ is the radiated or scattered power by the nanoparticle. The extinction cross section $C_{\textrm{ext}}$ for a lossless reciprocal nanoparticle is identical to the scattering cross section $C_{\textrm{sca}}$, i.e. $C_{\textrm{ext}}=C_{\textrm{sca}}$ due to the fact that the absorption cross section is zero $C_{\textrm{abs}} =C_{\textrm{ext}}-C_{\textrm{sca}}=0$. This explains why the scattering cross sections are identical for the proposed nanoparticle when illuminated from opposite directions, i.e. $C_{\textrm{sca}}=C_{\textrm{sca}}^{+}=C_{\textrm{sca}}^{-}$ [see black solid lines in Figs.~\ref{fig:figSCS}(a) and \ref{fig:figSCS}(c)]. Notice, the equality between extinction and scattering cross sections for the lossless bianisotropic nanoparticle leads to an expression known as the Sipe$-$Kranendonk condition\cite{Sipe:74, Tretyakov:03, Belov:03, Sersic:12}. It is important to note that for a plasmonic bianisotropic nanoparticle, due to the intrinsic Ohmic losses of metals, the scattering and absorption cross sections will be different for froward and backward illuminations~\cite{Sounas:14,Alaee:15}. Furthermore, a planar periodic array of lossy bianisotropic nanoparticles, possesses interesting optical features such as strongly asymmetric reflectance and perfect absorption. These effects have been studied before\cite{Menzel:10,Alaee:15}.
\begin{figure}
\centering
\includegraphics[width=0.49\textwidth]{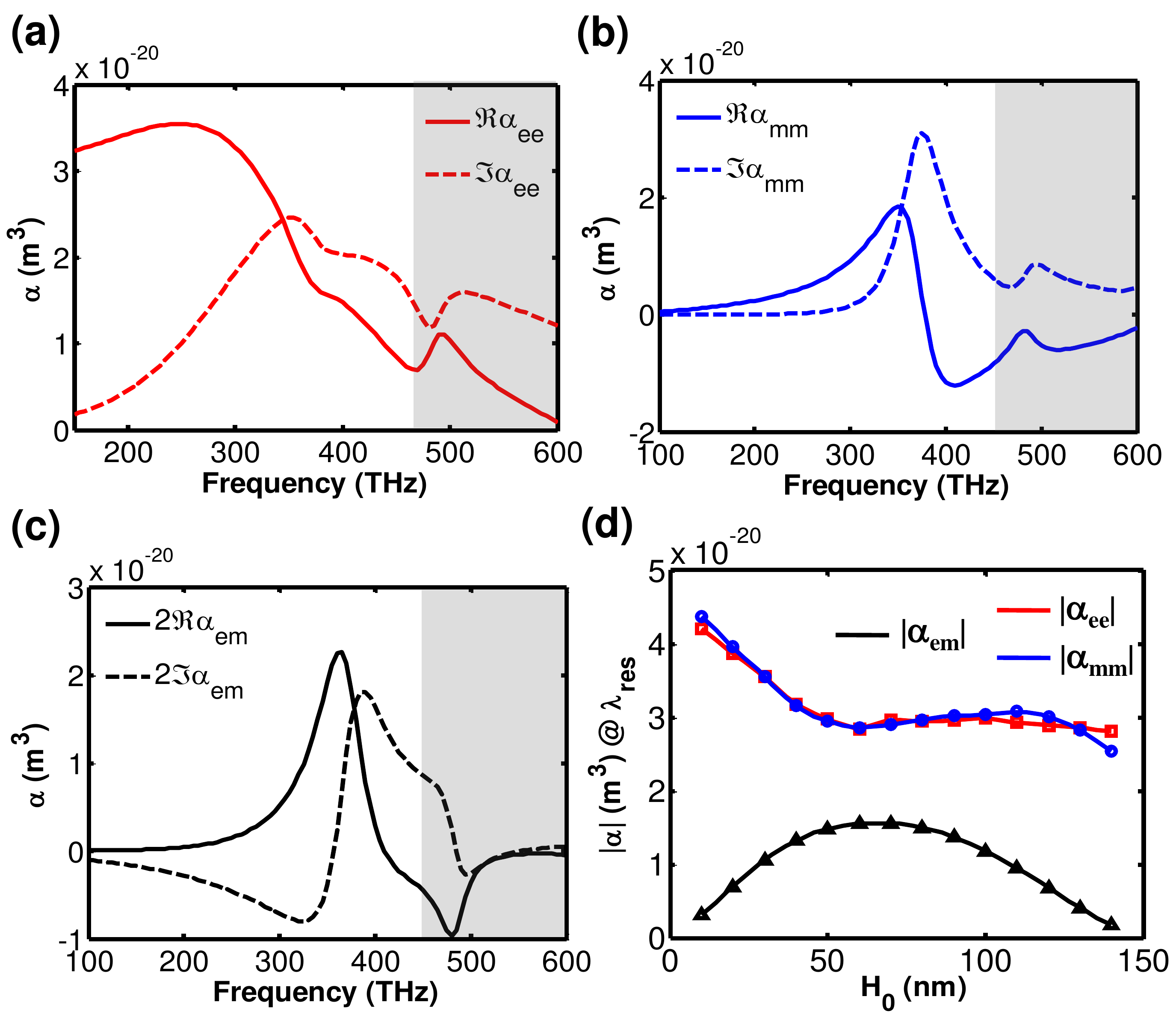}
\caption{(a)-(c) Individual polarizability components of the proposed nanoparticle. The geometrical parameters are: the height of the cylinder $H = 150$ nm, the height of cylindrical air hole $H_{0} = 100$ nm, and its diameter $D_{0} = 150$ nm. (d) Amplitude of the individual polarizability components at maximum of the magneto-electric coupling as a function of $H_{0}$. The maximum bianisotropy happens when the height of the air hole is approximately half of the height of the dielectric cylinder, i.e. $H_{0}\approx\frac{H}{2}=75$~nm.
}\label{fig:figAlpha}
\end{figure}
Although the scattering/extinction cross sections are similar for different illumination directions in the proposed nanoparticle, the contributions of the electric and magnetic dipole moments to the total scattering cross sections are not, i.e. $p_x^+\neq p_x^-$ and $m_y^+\neq m_y^-$  [see red and blue dashed lines in Figs.~\ref{fig:figSCS}(a) and \ref{fig:figSCS}(c)]. It can also be seen from Figs.~\ref{fig:figSCS}(b) and \ref{fig:figSCS}(d) that the magnetic field distributions are significantly different when the nanoparticle is illuminated from froward and backward directions. This is obviously due to the presence of magneto-electric coupling, which is introduced by the asymmetry in the nanoparticle.

In order to prove this, we have also calculated the individual polarizability components of the nanoparticle [see Figs.~\ref{fig:figAlpha} (a)-(c)] \cite{muhlig:11,Terekhov:11,Arango:14,Alaee:15}.
The magneto-electric coupling $\alpha_{\textrm{em}}$ is comparable with the electric $\alpha_{\textrm{ee}}$ and magnetic $\alpha_{\textrm{mm}}$ couplings for the proposed nanoparticle [see Figs.~\ref{fig:figAlpha} (a)-(c)]. The level of this coupling can be tuned by changing the geometrical parameters of the nanoparticle, i.e. the height $H_0$ or the diameter $D_0$ of the partially drilled cylindrical air hole inside the high-index dielectric cylinder. Figure~\ref{fig:figAlpha}(d) presents the magnitude of electric, magnetic, and magneto-electric polarizabilities of the nanoparticle at the maximum of magneto-electric coupling ($|\alpha_{\textrm{em}}|^{\textrm{max}}$) as a function of the height of cylindrical air hole $H_0$. It confirms that the amplitude of the magneto-electric coupling can be tuned and its maximum occurs when the height of the hole $H_0$ is appropriately half the height of the dielectric cylinder $H$, i.e. $H_{0}\approx\frac{H}{2}=75$~nm. Notice, the magneto-electric coupling can also be tuned by changing the diameter of cylindrical air hole $D_0$ (not shown for the sake of brevity).

\begin{figure*}
\centering
\includegraphics[width=0.89\textwidth]{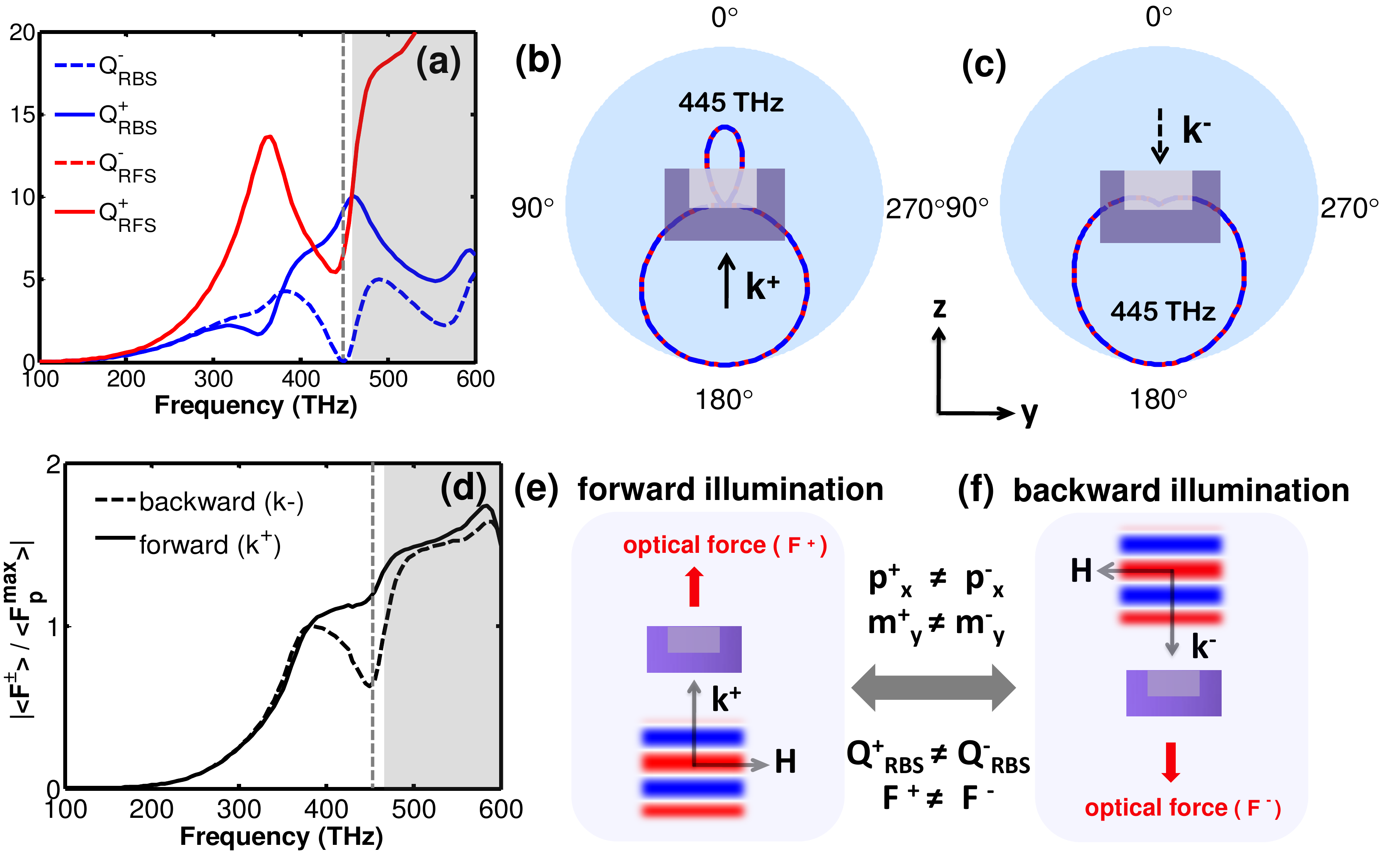}
\caption{(a) Normalized backward (blue lines) and forward (red lines) radar cross sections when the nanoparticle is illuminated from backward (dashed lines) and forward (solid lines) directions for $H_0=75$ nm. Note that the forward radar cross sections are identical and hence indistinguishable in the figure. (b) Radiation patterns for forward and (c) backward illumination directions at $445\,$THz, respectively. (d) Normalized forward (solid lines) and backward (dashed lines) forces. $F_p^{\textrm{max}} = \frac{3\pi}{k^{2}}\epsilon_{0}|\mathbf{E}_{\textrm{\textrm{i}nc}}|^{2}$ is the maximum force applied on a particle which exhibits an electric dipole moment with an incidence plane wave. (e) Schematic view of the different illuminations for forward and (f) backward illumination directions.}\label{fig:figQRSC}
\end{figure*}

\begin{figure*}
\centering
\includegraphics[width=0.85\textwidth]{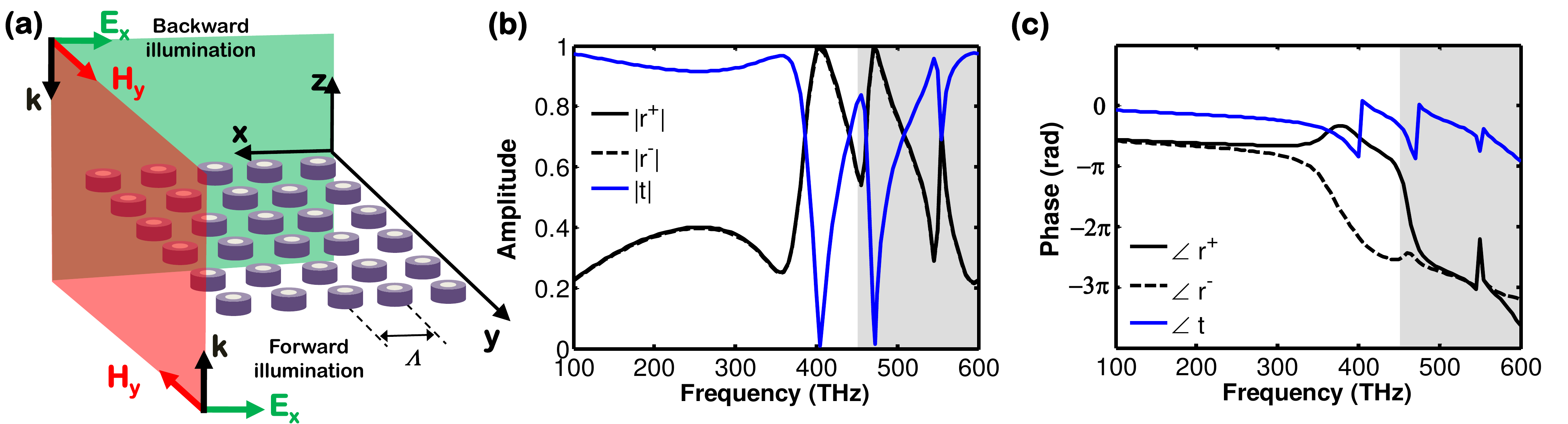}
\caption{(a) A metasurface composed of an array of bianisotropic nanoparticles with  $H_0=75$ nm. (b) Reflection and transmission amplitude for two illumination directions shown in (a). (c) The same plots as in (b) but for the corresponding phases. The reference plane for phase is considered at $z= 0$ plane.
}\label{fig:figarr}
\end{figure*}

As highlighted before, a lossless bianisotropic nanoparticle possesses identical scattering cross sections for both forward and backward illumination directions. Moreover, the forward radar cross section as shown in Fig.~\ref{fig:figQRSC}(a) are identical for both illumination directions according to the definition of the normalized forward radar cross section\cite{Alaee:15}:
\begin{eqnarray}\label{Q_RFS}
Q_{\textrm{RFS}}^{\pm} & = & \underset{r\rightarrow\infty}{\textrm{lim}}\frac{4\pi r^{2}}{A}\frac{|\mathbf{E}_{\textrm{sca}}\left(\varphi=0,\theta^{\pm}=0,\pi\right)|^{2}}{|\mathbf{E}_{\textrm{inc}}|^{2}}\nonumber\\
 & = & \frac{k^{4}}{4\pi A}\left|\alpha_{\textrm{ee}}\pm\alpha_{\textrm{em}}\pm\alpha_{\textrm{me}}+\alpha_{\textrm{mm}}\right|^{2}\nonumber\\
 & = & \frac{k^{4}}{4\pi A}\left|\alpha_{\textrm{ee}}+\alpha_{\textrm{mm}}\right|^{2}.
\end{eqnarray}
However, the backward radar cross section depends on the illumination directions [see Fig. \ref{fig:figQRSC}(a)]. It measures the share of light that is directly back reflected. In order to show that, we start with the definition of normalized backward radar cross section for both illuminations $Q_{\textrm{RBS}}^{\pm}$, i.e. backward radar cross section divided to the geometrical cross section \cite{Alaee:15} \begin{eqnarray}\label{Q_RBS1}
Q_{\textrm{RBS}}^{\pm} & = & \underset{r\rightarrow\infty}{\textrm{lim}}\frac{4\pi r^{2}}{A}\frac{|\mathbf{E}_{\textrm{sca}}\left(\varphi=0,\theta^{\pm}=\pi,0\right)|^{2}}{|\mathbf{E}_{\textrm{inc}}|^{2}}\nonumber\\
 & = & \frac{k^{4}}{4\pi A}\left|\alpha_{\textrm{ee}}\mp\alpha_{\textrm{em}}\pm\alpha_{\textrm{me}}-\alpha_{\textrm{mm}}\right|^{2},
\end{eqnarray}
where $A=\pi\left(D/2\right)^{2}$ is the geometrical cross section and $D$ is the diameter of the nanoparticle [see Fig.~\ref{fig:figGeo}]. Therefore, for the reciprocal nanoparticle, i.e., $\alpha_{\textrm{em}}=-\alpha_{\textrm{me}}$, the normalized backward radar cross section $Q_{\textrm{RBS}}^{\pm}$ can be written as \cite{Alaee:15} \begin{eqnarray}\label{Q_RBS2}
Q_{\textrm{RBS}}^{\pm} & = & \frac{k^{4}}{4\pi A}\left|\alpha_{\textrm{ee}}\mp2\alpha_{\textrm{em}}-\alpha_{\textrm{mm}}\right|^{2}.\label{eq:Q_RBS2}
\end{eqnarray}
Consequently, due to bianisotropy, the backscattering responses $Q_{\textrm{RBS}}^{\pm}$ are significantly different when the nanoparticle is illuminated from opposite directions [see Fig.~\ref{fig:figQRSC}(a)]. Hence, the nanoparticle possesses different radiation patterns for both illumination directions as depicted in Figs.~\ref{fig:figQRSC}(b) and \ref{fig:figQRSC}(c).

The dominant effect of the backscattering on optical scattering forces is shown for a dipolar particle without a magneto-electric response \cite{Nieto-Vesperinas:11}. Since, the investigated bianisotropic nanoparticle shows a notable backscatterig difference, it is interesting to explore the optical scattering forces in the realm of the bianisotropic particles.

The optical scattering force on a dipolar particle for a plane wave incident field reads as \cite{Chen:11}
\begin{eqnarray}\label{FR}
<\mathbf{F}> & = & \frac{1}{2}\textrm{Re}\left(\nabla\mathbf{E}_{\textrm{inc}}^{*}\cdot\mathbf{p}\right)+\frac{1}{2}\mu_{0}\textrm{Re}\left(\nabla\mathbf{H}_{\textrm{inc}}^{*}\cdot\mathbf{m}\right)\nonumber\\
 &  & -\frac{k^{4}}{12\pi\epsilon_{0}c}\textrm{Re}\left(\mathbf{p}\times\mathbf{m^{*}}\right)
\end{eqnarray}
Plugging Eqs.~(\ref{pm1}) and ~(\ref{pm2}) into Eq.~(\ref{FR}), one drives the optical scattering force for both illuminations:
\begin{eqnarray}\label{FR_Bia}
<\mathbf{F}^{\pm}> & = & \pm\mathbf{e}_{z}\frac{k}{2}\epsilon_{0}|E_{\textrm{x}}^{\textrm{\textrm{i}nc}}|^{2}\left\{ \textrm{Im}\left(\alpha_{\textrm{ee}}\pm\alpha_{\textrm{em}}\right)\right.\\
 &  & \pm\textrm{Im}\left(\alpha_{\textrm{me}}\pm\alpha_{\textrm{mm}}\right)\nonumber\\
 &  & \left.\mp\frac{k^{3}}{6\pi}\textrm{Re}\left[\left(\alpha_{\textrm{ee}}\pm\alpha_{\textrm{em}}\right)\left(\alpha_{\textrm{me}}\pm\alpha_{\textrm{mm}}\right)^{*}\right]\right\}.\nonumber
\end{eqnarray}
The difference between the forces applied on the nanoparticle in the direction of propagation assuming reciprocal nanoparticle leads to
\begin{eqnarray}\label{FR_Bia_2}
\bigtriangleup F & = & |<\mathbf{F}^{+}>|-|<\mathbf{F}^{-}>|\\
 & = & \frac{k^{4}}{6\pi}\epsilon_{0}|E_{\textrm{x}}^{\textrm{\textrm{i}nc}}|^{2}\textrm{Re}\left(\alpha_{\textrm{em}}\alpha_{\textrm{mm}}^{*}-\alpha_{\textrm{ee}}\alpha_{\textrm{em}}^{*}\right).\nonumber
\end{eqnarray}
Equation~(\ref{FR_Bia_2}) shows that the magneto-electric polarizability can lead to different optical scattering forces for opposite illuminations. Figure~\ref{fig:figQRSC} (d) shows the normalized optical scattering forces. It can be seen that $\bigtriangleup F$ is pronounced when the nanoparticle exhibits a notable difference in backward radar cross sections (i.e. at 445 THz).

In summary, the following relations hold for the proposed lossless bianisotropic nanoparticle:\begin{eqnarray}\label{Rels}
p^+ & \neq & p^-, \quad m^+  \neq  m^-,\nonumber
\\
C_{\textrm{sca}} & = &  C_{\textrm{ext}} = C_{\textrm{scat}}^{\pm} = C_{\textrm{ext}}^{\pm}, \quad Q_{\textrm{RFS}}^+  =  Q_{\textrm{RFS}}^-,\nonumber \\
\quad Q_{\textrm{RBS}}^+ &\neq&  Q_{\textrm{RBS}}^-,\quad |<\mathbf{F}^{+}>| \neq  |<\mathbf{F}^{-}>|.
\end{eqnarray}
Now we finish the investigations of the properties of the individual nanoparticle and start to demonstrate its interesting characteristics when used in a planar periodic array.

\section{Bianisotropic metasurfaces}

Next, we consider a periodic array composed of the proposed bianisotropic nanoparticle that is arranged along a planar surface, here the $xy$-plane [see Fig.~\ref{fig:figarr}(a)]. Since the nanoparticles are assumed to be lossless, the Ohmic losses of the proposed array is also zero. Let us illuminate this array at normal incidence from two directions, i.e. forward and backward direction [see Fig.~\ref{fig:figarr}(a) for the definition of forward and backward illumination]. The reflection and transmission amplitudes and phases for these two illumination cases are plotted in Figs.~\ref{fig:figarr}(b) and \ref{fig:figarr}(c). They were obtained from full wave simulations using the COMSOL Multiphysics \cite{multiphysics2012}. It is obvious that the amplitudes for the reflection and transmission for the different illumination directions are identical since there are no losses. The phase in transmission is equally identical due to reciprocity. On the contrary, the phase in reflection is not. This is apparently due to the presence of bianisotropy in the proposed array.

\begin{figure}
\centering
\includegraphics[width=0.49\textwidth]{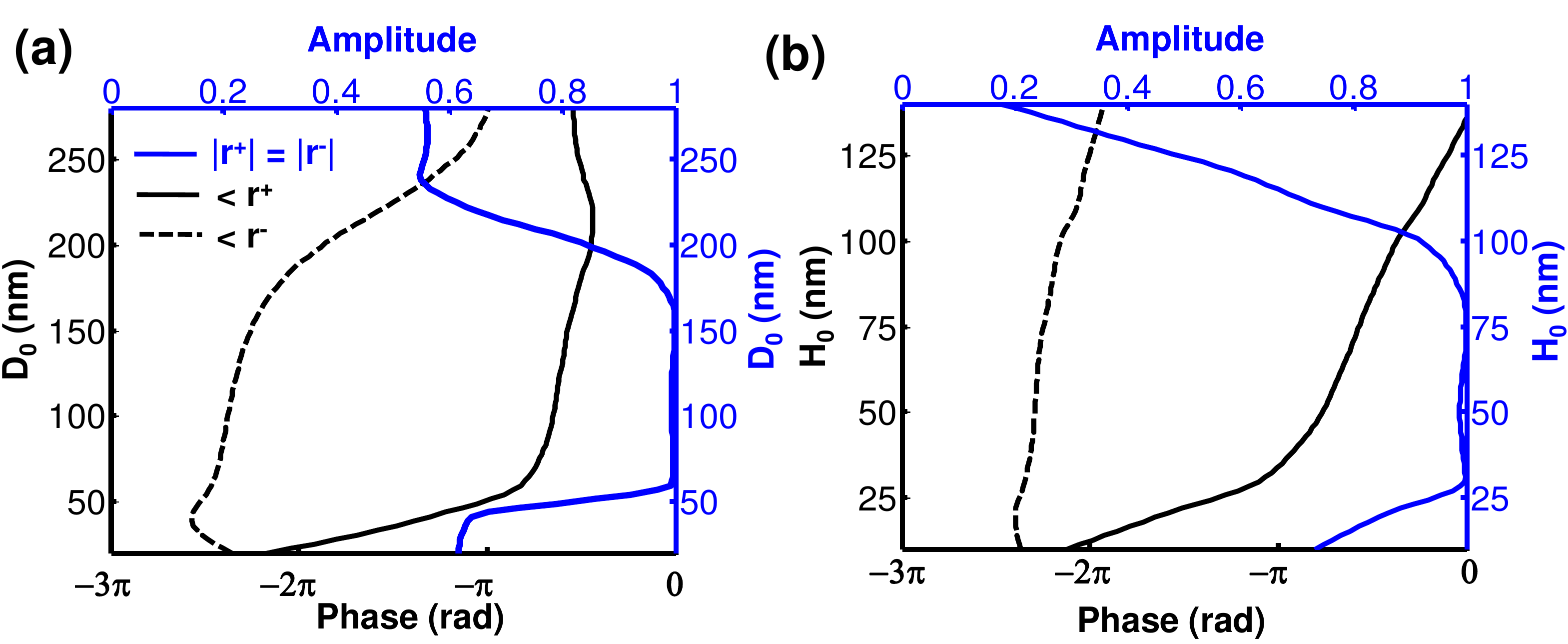}
\caption{(a) Phase-change curve for the reflection coefficient of an array of nanoparticles versus the diameter of air hole at resonance frequency $f=405$~THz for two different illumination directions denoted by $r^+$ and $r^-$. $H_0$ is fixed to $75$~nm. (b) The same plot as in (a) for variation of air hole height. $D_0$ is fixed to $150$~nm.
}\label{fig:figphas}
\end{figure}

Now, we claim that the proposed nanoparticle is a proper candidate as a unit cell in a reflectarrays. Indeed, a unit cell must provide a $2\pi$ phase change prior being applicable in a reflectarray \cite{Huang:05,Albooyeh:09,Kildishev:13,Cheng:14,Decker:15}. This $2\pi$ phase change shall be obtained by a suitable geometrical tuning. Moreover, it has to maintain a high reflection amplitude across the considered geometrical configuration. Figure ~\ref{fig:figphas} shows the amplitude and phase variations versus different height and diameter of the air hole inside the dielectric cylinder when the proposed nanoparticle is used in a square array with period $\Lambda=400$~nm.

As can be seen from Fig.~\ref{fig:figphas}(a), the proposed nanoparticle satisfies the required condition for the phase variation of the reflection coefficient for both illumination directions when we fix the height $H_0 = 75$ nm and change the diameter $D_0 = 20-280$ nm. The amplitude of the reflection coefficient is close to unity within an acceptable range while it drops down to one half at both ends of the diameter variations. The situation is a bit worse for the case when we fix the diameter $D_0 = 150$ nm and vary the height $H_0 = 10-140$ nm. In this case, we obtain a phase variation of $2\pi$ only for the forward illumination direction while the amplitude of the reflection coefficient, in some parameter regions, drops down to $20\%$ [see Fig.~\ref{fig:figphas}(b)]. We should mention that better results might be obtained if the geometry is carefully optimized. The presented example only serves the purpose to demonstrate the idea.

Another important point is that the proposed nanoparticles in the array, provides asymmetric phase variations. That is, the variation of reflection phases are different for different illumination directions. It means that we may properly design a reflectarray with two different properties when looking from different directions onto the plane. This is impossible using symmetrical structures which do not provide bianisotropic properties.

The most important point about the proposed reflectarray is its transparency out of the resonance band. Indeed, in the investigated reflectarray, we do not rely on a metallic back reflector \cite{Kildishev:13,Farmahini-Farahani:13,Yu:14}. 
Instead, we have offered resonant nanoparticles with bianisotropic properties to obtain a full reflection. This gives us the possibility to preserve transparency outside the resonance band in the reflectarray. This transparency is very important when combining multiple receiving/transmitting systems. Then, we need a reflectarray at a frequency band while we do not want to prevent signals to get transmitted out of that frequency band.


\section{Conclusions}

We have proposed a novel design for high-index dielectric nanoparticle which supports an omega-type bianisotropic coupling in addition to magnetic and electric optically-induced dipole resonances. We have demonstrated that the magneto-electric coupling can be tuned by geometric parameters of the nanoparticle. In particular, the nanoparticle can possess different backward cross sections and consequently different optical scattering forces, when being illuminated by a plane wave from opposite directions along with the identical forward cross sections.

For a metasurface created by a periodic array of bianisotropic nanoparticles, we have observed interesting effects, e.g. asymmetric reflection phases for the opposite illumination directions, and a possibility to achieve a $2\pi$ phase change together with an acceptable reflection amplitude across the entire phase spectrum by tuning the geometric parameters of the nanoparticle. Finally, we have demonstrated that the employment of the proposed resonant nanoparticle together with the absence of a fully reflective metallic screen gives the opportunity of \textit{out-of-band transparency} in reflectarrays.

All the effects described here are the direct outcome of the engineered omega-type bianisotropy that may open a new direction to design high-index nanoparticles and metasurfaces, including reflectarrays, transmitarrays, and Huygens metasurfaces.

\section*{ Acknowledgements}

The authors thanks the German Science Foundation (project RO 3640/7-1) for a financial support. The work was also supported by the DAAD (PPP Australien) and the Australian Research Council.
\%

\end{document}